\documentclass{osa-article}

\journal{oe}
\articletype{Research Article}
\usepackage{lineno}
\begin{document}

\title{ All-optical pulse switching with a periodically driven dissipative quantum system}

\author{Yingying Han,\authormark{1} Wenxian Zhang,\authormark{2} and Weidong Li\authormark{1,3,*}}

\address{\authormark{1}Shenzhen Key Laboratory of Ultraintense Laser and Advanced Material Technology, Center for Advanced Material Diagnostic Technology, and College of Engineering Physics, Shenzhen Technology University, Shenzhen 518118, China\\
\authormark{2}Key Laboratory of Artificial Micro- and Nano-structures of Ministry of Education, and School of Physics and Technology, Wuhan University, Wuhan, Hubei 430072, China\\
\authormark{3}Guangdong Provincial Key Laoratory of Quantum Science and Engineering, Southern University of Science and Technology, Shenzhen 518055,China}

\email{\authormark{*}liweidong@sztu.edu.cn} %% email address is required



\begin{abstract}
All-optical switching used to switch the input optical signals without any electro-optical conversion plays a vital role in the next generation of optical information processing devices. Even all-optical switchings (AOSs) with continuous input signals have been widely studied, all-optical pulse switchings (AOPSs) whose input signals are pulse sequences have rarely been investigated because of the time-dependent Hamiltonian, especially for dissipative quantum systems. In this paper, we propose an AOPS scheme, where a strong pulsed field is used to switch another pulsed input signal. With the help of Floquet-Lindblad theory, we identify the control field that can effectively turn on/off the input signal whose amplitude envelope is a square-wave (SW) pulse train in a three-level dissipative system. By comparing the properties of the AOPSs controlled by a continuous-wave (CW) field and an SW control field, we find that the SW field is more suitable to be a practical tool for controlling the input SW signal. It is interesting to impress that the switching efficacy is robust against pulse errors. The proposed protocol is readily implemented in atomic gases or superconducting circuits, and corresponds to AOPSs or all-microwave pulse switchings.
\end{abstract}

%%%%%%%%%%%%%%%%%%%%%%%%%%  body  %%%%%%%%%%%%%%%%%%%%%%%%%%
\section{Introduction}
All-optical signal processing, which processes optical signals without converting to electrical form, has always been critical for the development of fast optical networks~\cite{Minzioni_2019}. In order to realize all-optical signal processing, several types of all-optical devices are required, including all-optical modulators~\cite{MXene2020,Yuan:21}, all-optical flip-flops~\cite{An2010,IST2005,Optical2010}, all-optical logic gates~\cite{Ultrafast,All2012,Optical2005}, and AOSs~\cite{All2006,Almeida, 1992All}.

All-optical switchings, in which the switching is produced by using a strong control field to control the propagation
(or scattering) of an input probe field, have great potential for use in optical computation and communication~\cite{Ultrafast}. Most existing theoretical studies are based on the continuous-wave (CW) model that assumes continuous monochromatic input fields~\cite{Boyraz,LiDynamical}, but practical applications and experimental demonstrations of switching usually employ pulses~\cite{Shaped,Malinovskaya,Parmigiani,Ultrafast2007}. For a few specific single (not pulse train) pulses, some numerical simulations have been carried out~\cite{PhysRevLett.68.911, Ramos,PhysRevLett.95.043001,NOVITSKY2016202}. However, to the best of our knowledge, the studies on AOPS are rare, especially the analytic solutions, because the time-dependent Floquet-Lindblad equation makes them a challenging task.

In this work, we present a theory of the AOPS for an input field whose amplitude envelope is an SW sequence in a three-level dissipative system. A variety of effects can be employed for switching~\cite{BirnbaumPhoton, Dawes2005All, PhysRevA.68.041801, Wang2002Controlling}. The one we applied here is the transparency window phenomena\cite{Chen:05, PhysRevLett.102.203902}. The on (off) state is realized when the control field intensity is sufficiently strong (zero) to generate a transparency window (adsorption peak), suppressing (increasing) the absorption of the input field and resulting in a maximum (minimum) of transmission~\cite{Stern2017Strong}. We obtain the analytical results of periodically driven dissipation quantum systems under high-frequency expansions by using Floquet-Lindblad theory and Van Vleck perturbation theory~\cite{PhysRevB.104.165414,2020General,PhysRevA.101.022108}, which agree well with the numerical results. Further comparing the properties of the AOPSs controlled by a CW field and an SW field, we find that the SW field is a proper control field for the input pulse we assumed here. Finally, we designed an AOPS, in which a strong pulsed field is used to turn on and off the pulsed input field.

The paper is organized as follows. In Sec.~\ref{sec:mod}, we describe the model: a dissipation quantum system interacting with a pulsed input field. In Sec.~\ref{sec:ana}, we investigate the analytical results of periodically driven dissipation quantum systems under high-frequency expansions by using Floquet-Lindblad theory and Van Vleck perturbation theory. In Sec.~\ref{sec:allo}, we compare the properties of the AOPSs controlled by a strong CW field and a strong SW field. Finally, the conclusion is given in Sec.~\ref{sec:con}.

\section{A dissipative quantum system interacting with a pulsed input field}
\label{sec:mod}

\begin{figure}[thp]
\centering
\includegraphics[width=3.2in]{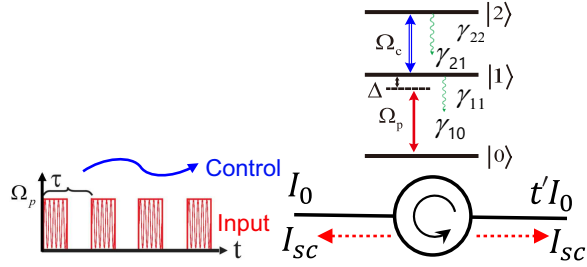}
\caption{\label{fig:mod} A three-level system denoted by $|0\rangle,~|1\rangle$ and $|2\rangle$ with energy $\omega_{0,1,2}$ couples with electromagnetic fields confined in a 1D transmission line. An input field with amplitude $I_0$ couples states $|0\rangle$ and $|1\rangle$ with a detuning $\Delta=\omega_{1}-\omega_0-\omega_{p}$ and the transmission coefficient is $t'$. $I_{sc}$ is the amplitude of the scattering wave, propagating in both directions (forward and backward). The time dependence of the amplitude of the input field is approximated by an SW envelope with period $\tau$ and a duty cycle of 50$\%$. A strong control field resonantly couples states $|1\rangle$ and $|2\rangle$. $\gamma_{21}~(\gamma_{10})$ is the population damping rate from state $|2\rangle~(|1\rangle)$ to $|1\rangle~(|0\rangle)$. $\gamma_{22}~(\gamma_{11}$ is the dephasing rate of
 state $|2\rangle~(|1\rangle)$.}
\end{figure}
We consider a three-level atom interacting with a field of an electromagnetic 1D wave~\cite{PhysRevLett.95.213001, Resonance}. The amplitude of the input probe field is modulated by an SW envelope with period $\tau$, as shown in Fig.~\ref{fig:mod}. In the semiclassical theory of quantum optics, the field of the input wave $I_0(x,t)=I_0(t)e^{ikx-i\omega_p t}$ (where $\omega_p$ is the frequency and k is the wave number) induces the atomic polarization. The atom placed at $x=0$ generates waves $I_{sc}(x,t)=I_{sc}(t)e^{ik|x|-i\omega_p t}$, propagating in forward or backward direction. Here we assume the general scenario in which the atom size is negligibly small as compared with the wavelength~\cite{Resonance} and investigate elastic Rayleigh scattering in which the incident and the scattered waves have the same frequency. It is convenient to define reflection and transmission coefficients $r$ and $t'$ according to $I_{sc}(t)=-rI_0(t)$ and $I_0(t)+I_{sc}(t)=t'I_0(t)$, therefore, $t'+r=1$. And $r$ is proportional to the off-diagonal element $\rho_{10}=\text{Tr}[\sigma_{10}\rho$] with transition operator $\sigma_{10}=|1\rangle\langle 0|$ and density matrix $\rho$\cite{PhysRevLett.104.193601,Resonance},
\begin{equation}\label{eq:reco}
r=\eta\frac{i\gamma_{10}}{\Omega_p}\rho_{10},
\end{equation}
where $\eta$ is a constant determined by experimental device and $\Omega_p$ is the Rabi-frequency of the input field.

For a traditional AOS~\cite{PhysRevLett.107.073601,2018All}, a strong CW control field is used to switch on/off the transmission of the input CW probe field. With the control off, the input is reflected from the transition line, and when the control is on, the input is transmitted through the transmission line and the reflection coefficient is almost zero (i.e., $r\approx0$). Inspired by these, to switch on/off the transmission of the pulsed input field, we introduce another auxiliary level $|2\rangle$ and a strong control field with frequency $\omega_c=\omega_{2}-\omega_{1}$. However, the characteristics of the control field (such as, continuous or pulsed field) are still unclear, and further research is needed. In the following, we will explore this problem in detail with both numerical and analytical methods by using Floquet-Lindblad theory. Here we take the ladder-type three-level system as an example. In fact, it can also be $\Lambda$-type or $V$-type, which is common in atomic gases~\cite{PhysRevA.64.041801}.

\section{Analytical results of periodically driven dissipative quantum systems}\label{sec:ana}

In this section, we investigate the analytical results of periodically driven dissipative quantum systems under high-frequency expansions by using Floquet-Lindblad theory and Van Vleck perturbation theory~\cite{PhysRevB.104.165414}.

\subsection{Formula}\label{sec:form}

Firstly, we introduce the general description for the nonequilibrium steady states (NESSs), which are the long-time dynamics of periodically driven dissipative quantum systems. For a quantum system defined on an N-dimensional Hilbert space, $H_0$ denote the time-independent Hamiltonian with eigenenergies $\{E_i\}_{i=1}^N$ and eigenstates $\{|E_i\rangle\}_{i=1}^N$. Here $E_1<E_2<\cdots<E_N$. The effect of the driving is represented by a time-dependent Hamiltonian $H_{\text{ext}}(t)$ with period $\tau$: $H_{\text{ext}}(t+\tau)=H_{\text{ext}}(t)$. The total Hamiltonian is $H(t)=H_0+H_{\text{ext}}(t)$ and hence $H(t+\tau)=H(t)$. Thus the Fourier series of $H(t)$ can be written as~\cite{PhysRevA.79.032301}
\begin{equation}\label{eq:fh}
H(t)=\sum_{n=-\infty}^{\infty}H_n e^{-i n\omega t}
\end{equation}
with Floquet frequency $\omega=2\pi/\tau$.

To study dissipative systems, we consider the density operator $\rho(t)$, within the Born-Markov approximation~\cite{Quantum}, whose dynamics is described by the Floquet-Lindblad equation~\cite{2020General,PhysRevLett.127.070402,PhysRevA.93.032121} ($\hbar=1$ throughout this work)
\begin{eqnarray}\label{eq:tma}
  \partial_t\rho(t) &=& \mathcal{L}\rho(t)=-i[H(t),\rho(t)]+\mathcal{D}[\rho(t)], \\
  \mathcal{D}[\rho(t)] &\equiv& \sum_{i,j}\gamma_{ij} \left[L_{ij}\rho(t)L^{\dag}_{ij}-\frac{1}{2}\{L_{ij}^{\dag}L_{ij},\rho(t)\}\right],
\end{eqnarray}
%Note that here we assume the dissipative rates are smaller than the energy gaps of $H_0$ to satisfy the Born-Markov approximation.
with $\{A,B\}=AB+BA$ and projection operator $L_{ij}\equiv|E_i\rangle\langle E_j|$. The long-time dynamic of Eq.~({\ref{eq:tma}}) is NESS, which emerges in a balance of the energy injection by the periodic driving and the energy dissipation. The general description for the NESS under high Floquet frequency is given by~\cite{PhysRevB.104.165414}
\begin{equation}\label{eq:trho}
\rho(t)=e^{\mathcal{G}(t)}e^{t\mathcal{L}_{\text{eff}}}e^{-\mathcal{G}(0)}\rho(0).
\end{equation}
The time-dependent part $e^{\mathcal{G}(t)}$ is the micromotion operator periodic in time $\mathcal{G}(t+T)=\mathcal{G}(t)$.
On the basis of the high-frequency expansion, i.e., the Floquet frequency $\omega$ is greater than the energy scales of $H_0$, the micromotion operator is given by
\begin{equation}\label{eq:micro}
\mathcal{G}(t)\rho=\frac{1}{\omega}\sum_{m\neq0}[H_m,\rho]e^{-im\omega t}+O(\omega^{-2}).
\end{equation}
The time-independent part $\mathcal{L}_{\text{eff}}$ is represented by the effective Hamiltonian
\begin{equation}\label{eq:effl}
\mathcal{L}_{\text{eff}}(\rho)=-i[H_{\text{eff}},\rho]+\mathcal{D}(\rho)+O(\omega^{-2})
\end{equation}
with
\begin{equation}\label{eq:effh}
H_{\text{eff}}=H_0+\frac{1}{2}\sum_{n>0}\frac{[H_{-n},H_n]}{n}+O(\omega^{-2}).
\end{equation}
Focusing on the leading-order contribution, Eq.~({\ref{eq:trho}) has a simple explicit formula~\cite{2020General}
\begin{equation}\label{eq:strho}
\rho_{\text{ness}}(t)=\tilde{\rho}+\sigma_{\text{MM}}(t)+\sigma_{\text{FE}}+O(\omega^{-2}),
\end{equation}
in which both $\sigma_{\text{MM}}(t)$ and $\sigma_{\text{FE}}$ are $O(\omega^{-1})$. $\sigma_{\text{MM}}(t)$
and $\sigma_{\text{FE}}$ are the micromotion and Floquet-engineering parts, respectively. The micromotion part
$\sigma_{\text{MM}}(t)$ is
\begin{equation}\label{eq:tmocro}
\sigma_{\text{MM}}(t)=\frac{1}{\omega}\sum_{m\neq0}\frac{e^{-im\omega t}}{m}[H_m,\tilde{\rho}],
\end{equation}
where $\tilde{\rho}$ is characterized by
\begin{equation}\label{eq:rhot}
-i[H_0,\tilde{\rho}]+\mathcal{D}(\tilde{\rho})=\dot{\tilde{\rho}}=0.
\end{equation}
There are two properties of $\sigma_{\text{MM}}(t)$: (i) $\sigma_{\text{MM}}(t+\tau)=\sigma_{\text{MM}}(t)$ and it contributes to oscillations of physical observable, and (ii) $\sigma_{\text{MM}}(t)$ does not contribute to the time averages of physical observable for one period of oscillation.  The time-independent Floquet-engineering part $\sigma_{\text{FE}}$ is given by
\begin{equation}\label{eq:fet}
\langle E_k|\sigma_{\text{FE}}|E_l\rangle=\frac{\langle E_k|\Delta H_{\text{eff}}|E_l\rangle}{E_k-E_l-i\gamma_{kl}}
(\tilde{p}_k-\tilde{p}_l)~~~~(for k\neq l)
\end{equation}
and $\langle E_k|\sigma_{FE}|E_k\rangle=0$ for all $k$, where $\Delta H_{\text{eff}}\equiv H_{\text{eff}}-H_0=O(\omega^{-1})$,
$\tilde{p}_k=\langle E_k|\tilde{\rho}|E_k\rangle$ and $\gamma_{kl}=\sum_i(\gamma_{ik}+\gamma_{il})/2$. Floquet-engineering
part $\sigma_{\text{FE}}$ describes how the effective Hamiltonian changes physical observable from their values in static
counterpart and contributes to the time-averaged quantities. These formulas provide powerful tools for analyzing the NESSs of periodically driven dissipative quantum systems.

\subsection{Dissipative two-level systems}\label{sec:two}
\begin{figure}[thp]
\centering
\includegraphics[width=3.2in]{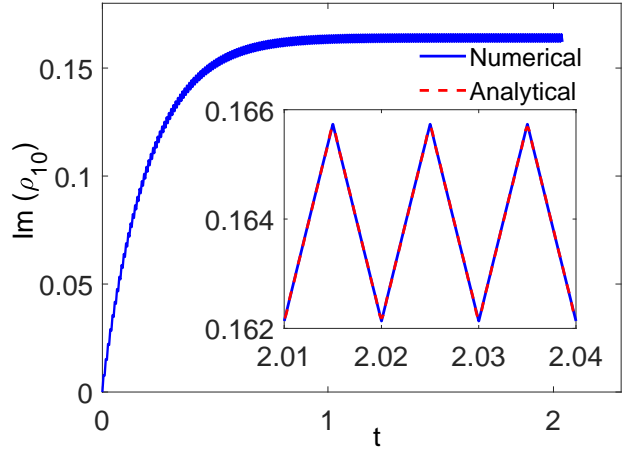}
\caption{\label{fig:two} Time evolutions of the off-diagonal element $\text{Im}(\rho_{10})$ for
$\tau=0.01, \Omega_p/(2\pi)=0.5$, and $\Delta=0$. The blue line denotes the numerical result calculated from the Floquet-Lindblad equation [Eq.~(\ref{eq:tma})] with time-dependent Hamiltonian Eq.~(\ref{eq:two}), a sufficiently short time step (here, $\delta t=0.00001$) and initial state $|\Psi(0)\rangle=|0\rangle$. Inset: comparison of the numerical and analytical results of $\text{Im}(\rho_{10})$, when the system reaches NESS. The red dashed line is the analytical result calculated from Eq.~(\ref{eq:nesstwo}). Clearly, the analytical and numerical results agree well with each other.}
\end{figure}
We now calculate the NESS of a dissipative two-level system with the control field off. As shown in Fig.~\ref{fig:mod}, a generic two-level system denoted by $|0\rangle,~|1\rangle$ driven by an SW-pulse input field with detuning $\Delta$ and time-dependent Rabi-frequency $\Omega_p(t+\tau)=\Omega_p(t)$. The Hamiltonian can be written as
\begin{eqnarray}\label{eq:two}
  H(t) = \frac{\Delta}{2}(|1\rangle\langle1|-|0\rangle\langle0|)
   - [\frac{\Omega_p(t)}{2}|0\rangle\langle1|+h.c.].
\end{eqnarray}
We have adopted the rotating wave approximation by assuming $\omega,~\Omega_p\ll \omega_p$, which is also
the parameter ranges of many experiments. For example, in superconducting circuit (atom) experiments, the
frequency of near-resonant microwave (laser) is several GHz (THz), however, the Floquet frequency and the Rabi
frequency of control and probe fields are about several or hundreds MHz~\cite{HanPRApplied}. Note that high-frequency approximation, i.e., $\omega>\Delta,\Omega_p$, do not break this condition.

Expand the square-wave function $\Omega_p(t)$, as shown in Fig.~\ref{fig:mod}, into many Fourier components:
\begin{eqnarray}\label{eq:omept}
\Omega_{p}(t)=\frac{\Omega_p}{2}+\sum_{n=1}^{\infty}\Omega_{pn}\sin(\omega_n t),
\end{eqnarray}
where
\begin{eqnarray}
  \Omega_{pn} &=& \frac{2\Omega_p}{(2n-1)\pi}, \\
  \omega_n &=& (2n-1)\omega,~~~n=1,2,3\cdots .
\end{eqnarray}
Substituting Eqs.~(\ref{eq:two}) and (\ref{eq:omept}) into Eq.~(\ref{eq:fh}), the Floquet matrix blocks are given by
\begin{eqnarray}
  H_0 &=& \frac{\Delta}{2}(|1\rangle\langle1|-|0\rangle\langle0|)- \left[\frac{\Omega_p}{4}|0\rangle\langle1|+h.c.\right],
  \label{eq:twoh0}\\
  H_{2n-1} &=& H^*_{-(2n-1)}=\frac{i\Omega_{pn}}{4}|0\rangle\langle1|+\frac{i\Omega_{pn}}{4}|1\rangle\langle0|, \label{eq:twoho}\\
  H_{2n} &=& H^*_{-2n}=0,\label{eq:twohe}
\end{eqnarray}
where we have used the function $\sin(\omega_n t)=i(e^{-i\omega_nt}-e^{i\omega_n t})/2$.
According to Eq.~(\ref{eq:strho}), to describe the NESS of this dissipative two-level system, we need to solve
Eq.~(\ref{eq:rhot}). Substituting Eq.~(\ref{eq:twoh0}) to Eq.~(\ref{eq:rhot}), we have
\begin{eqnarray}
  \partial_t\tilde{\rho}_{11} &=& -\gamma_{10}+\frac{i\Omega_p}{4}(\rho_{01}-\rho_{10}), \label{eq:wrho11}\\
  \partial_t\tilde{\rho}_{10} &=& -(i\Delta+\Gamma_1)\rho_{10}+\frac{i\Omega_p}{4}(\rho_{00}-\rho_{11}), \label{eq:wrho10}\\
  \partial_t\tilde{\rho}_{01} &=& (i\Delta-\Gamma_1)\rho_{01}- \frac{i\Omega_p}{4}(\rho_{00}-\rho_{11})\label{eq:wrho01}
\end{eqnarray}
with $\Gamma_1=\gamma_{10}/2+\gamma_{11}$. Combined with $\rho_{00}+\rho_{11}=1$, the steady solutions of
Eqs.~(\ref{eq:wrho11})-(\ref{eq:wrho01}) are
\begin{eqnarray}\label{eq:tsrho11}
  \tilde{\rho}_{11} &=& \frac{\Omega_p^2\Gamma_1/2}{4\gamma_{10}\Delta^2+4\Gamma_1^2\gamma_{10}+\Omega_p^2\Gamma_1}, \\
  \tilde{\rho}_{10} &=& \frac{(\Delta+i\Gamma_1)\Omega_p}{4\Delta^2+4\Gamma_1^2+\Omega_p^2\Gamma_1/\gamma_{10}}, \\
  \tilde{\rho}_{01}&=& \frac{(\Delta-i\Gamma_1)\Omega_p}{4\Delta^2+4\Gamma_1^2+\Omega_p^2\Gamma_1/\gamma_{10}}\label{eq:tsrho01}.
\end{eqnarray}

In the calculations of the AOPSs in this work, the input probe field resonates with the transition between $|0\rangle$ and $|1\rangle$ to obtain the best switching performance ~\cite{LiDynamical}, i.e., $\Delta=0$, then Eqs.~(\ref{eq:tsrho11})-(\ref{eq:tsrho01}) become
 \begin{eqnarray}\label{eq:rho11}
  \tilde{\rho}_{11} &=& \frac{\Omega_p^2\Gamma_1/2}{4\Gamma_1^2\gamma_{10}+\Omega_p^2\Gamma_1}, \\
  \tilde{\rho}_{10} &=& \frac{i\Gamma_1\Omega_p}{4\Gamma_1^2+\Omega_p^2\Gamma_1/\gamma_{10}}, \label{eq:rho10}\\
  \tilde{\rho}_{01}&=& \frac{-i\Gamma_1\Omega_p}{4\Gamma_1^2+\Omega_p^2\Gamma_1/\gamma_{10}}\label{eq:rho01}.
\end{eqnarray}
Substituting Eqs.~(\ref{eq:twoho})-(\ref{eq:twohe}) and (\ref{eq:rho11})-(\ref{eq:rho01}), into Eq.~(\ref{eq:tmocro}), the micromotion part $\sigma_{\text{MM}}(t)$ of $\rho_{10}(t)$ is given by
 \begin{equation}\label{eq:microt}
 \sigma_{\text{MM}}^{10}(t)=\frac{-4i\Gamma_1\gamma_{10}\Omega_p}{\pi\omega(4\Gamma_1\gamma_{10}+\Omega_p^2)}
 \sum_{n=1}^{\infty}\frac{\cos[(2n-1)\omega t]}{(2n-1)^2},
 \end{equation}
 which is a periodic triangular pulse with frequency $\omega$ and amplitude (difference between maximum and minimum)
 $\pi\Gamma_1\gamma_{10}\Omega_p/[\omega^2(4\Gamma_1\gamma_{10}+\Omega_p^2)]$. Note that in Eq.~(\ref{eq:microt}), we have omitted the higher order terms than $O(\omega^{-1})$, because the formula of $\sigma_{\text{MM}}^{10}(t)$ we used here is $O(\omega^{-1})$, as described in Eq.~(\ref{eq:strho}) and Eq.~(\ref{eq:tmocro}).

To calculate the Floquet engineering part $\sigma_{FE}$ in Eq.~(\ref{eq:fet}), we should get $\Delta H_{\text{eff}}$ and
\begin{equation}\label{eq:oo}
\Delta H_{\text{eff}}=\sum_{n>0}[H_{-n},H_n]/(2n)+O(\omega^{-2})=O(\omega^{-2}),
\end{equation}
since $H_{n}=-H_{-n}$. On the basis of high-frequency expansion (more precisely, $\omega>\Delta,\Omega_p$), we neglect higher order terms [i.e., higher than $O(\omega^{-1})$]. Therefore, focusing on the leading-order contribution, the NESS of $\rho_{10}(t)$ is given by
 \begin{eqnarray}\label{eq:nesstwo}
 \rho_{10}^{\text{ness}}(t)&=&\frac{i\Gamma_1\Omega_p}{4\Gamma_1^2+\Omega_p^2\Gamma_1/\gamma_{10}}
 -\frac{4i\Gamma_1\gamma_{10}\Omega_p}{\pi\omega(4\Gamma_1\gamma_{10}+\Omega_p^2)}
 \sum_{n=1}^{\infty}\frac{\cos[(2n-1)\omega t]}{(2n-1)^2},
 \end{eqnarray}
in which $\text{Re}(\rho_{10})=0$.

In Fig.~\ref{fig:two}, we present the dynamics of the off-diagonal element $\text{Im}(\rho_{10})$ with parameters $\tau=0.01$, $\Omega_p/(2\pi)=0.5$, $\Delta=0$, $\gamma_{10}/(2\pi)=1$ and $\gamma_{11}/(2\pi)=0.2$. We find that it oscillates with time and finally reaches NESS after a sufficiently long time. Moreover, in the inset of Fig.~\ref{fig:two}, we compare the numerical and analytical results of $\text{Im}(\rho_{10})$ and find that they agree well with each other. Indeed, the micromotion part of $\text{Im}(\rho_{10})$ is a periodic triangular pulse predicted by Eq.~(\ref{eq:microt}).

 \subsection{Dissipative Floquet three-level systems}\label{sec:three}

In this subsection, we investigate the other situation: the input is transmitted through the transmission line with a strong
control field on and the reflection coefficient should be as small as possible. Based on the above two-level system, we further consider another level $|2\rangle$ with energy
$\omega_2>\omega_1>\omega_0$, as shown in Fig.~\ref{fig:mod}. Physically, there are dissipative rates and we set
population damping rate $\gamma_{21}/(2\pi)=1.2$ and dephasing rate $\gamma_{22}/(2\pi)=0.2$. Other parameters are the same as in Sec.~\ref{sec:two}. Note that here $\gamma_{20}=0$, because the transition between levels $|0\rangle$ and $|2\rangle$ is forbidden. A control field with frequency $\omega_c$ couples levels $|1\rangle$ and $|2\rangle$ resonantly, i.e., $\omega_{2}-\omega_{1}=\omega_{c}$. In this work, we consider two scenarios: (i) CW control field with constant Rabi-frequency $\Omega_c$, and (ii) SW control field with time-dependent Rabi-frequency $\Omega_c (t)$.

 \subsubsection{Continuous-wave (CW) control field}\label{sec:cwth}
 \begin{figure}[thp]
 \centering
\includegraphics[width=3.2in]{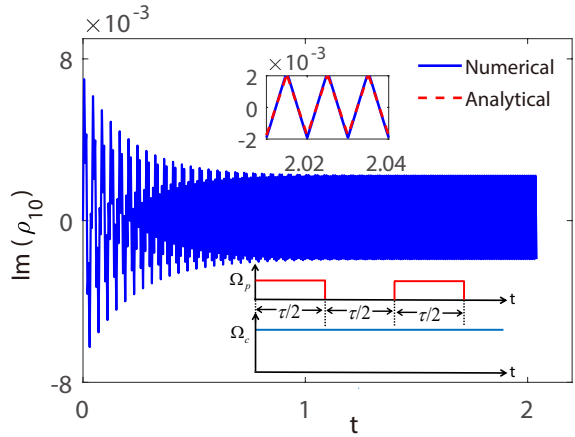}
\caption{\label{fig:cw} Time evolutions of the off-diagonal element $\text{Im}(\rho_{10})$ for $\tau=0.01$,
$\Omega_p/(2\pi)=0.5$, $\Omega_c/(2\pi)=50$, and $\Delta=0$. The blue line denotes the numerical result calculated from the Floquet-Lindblad equation [Eq.~(\ref{eq:tma})] with time-dependent Hamiltonian Eq.~(\ref{eq:threecw}) for sufficiently short time step $\delta t=0.00001$. Upper inset: comparison of the numerical and analytical results of $\text{Im}(\rho_{10})$, when the system reaches NESS. Clearly, the analytical and numerical results agree well with each other. The red dashed line is the analytical result calculated from Eq.~(\ref{eq:nesscw}). Lower inset: the amplitude envelops of the input and control fields.}
\end{figure}
 For the CW control field with constant Rabi-frequency, as the lower inset shown in Fig.~\ref{fig:cw}, the Hamiltonian
 of the system is
 \begin{eqnarray}\label{eq:threecw}
  H(t) &=& \frac{\Delta}{2}(|1\rangle\langle1|+|2\rangle\langle2|-|0\rangle\langle0|)
   - \frac{1}{2}[\Omega_p(t)|0\rangle\langle1|+\Omega_c|1\rangle\langle2|+h.c.],
\end{eqnarray}
with $\Omega_p(t)$ is given by Eq.~(\ref{eq:omept}). We have adopted the rotating wave approximation by assuming
$\omega, \Omega_p\ll \omega_p$, and $\Omega_c\ll\omega_c$. For superconducting circuit (atom) systems, $\omega_p/\gamma_{10}$ and $\omega_c/\gamma_{10}$ are several thousands (millions). Similarly, the Floquet matrix blocks are
\begin{eqnarray}
  H_0 &=& \frac{\Delta}{2}(|2\rangle\langle2|+|1\rangle\langle1|-|0\rangle\langle0|)
  -[\frac{\Omega_p}{4}|0\rangle\langle1|+\frac{\Omega_c}{2}|1\rangle\langle2|+h.c.], \label{eq:tcwh0}\\
  H_{2n-1} &=& H^*_{-(2n-1)}=\frac{i\Omega_{pn}}{4}|0\rangle\langle1|+\frac{i\Omega_{pn}}{4}|1\rangle\langle0|,\\
  H_{2n} &=& H^*_{-2n}=0.
\end{eqnarray}
Substituting Eq.~(\ref{eq:tcwh0}) to Eq.~(\ref{eq:rhot}) and with $\Delta=0$, we have
\begin{eqnarray}
  \partial_t\tilde{\rho}_{00} &=& \gamma_{10}\tilde{\rho}_{11}
  -\frac{i\Omega_p}{4}(\tilde{\rho}_{01}-\tilde{\rho}_{10}), \label{eq:cwdrho00}\\
  \partial_t\tilde{\rho}_{22} &=&-\gamma_{21}\tilde{\rho}_{22}
  +\frac{i\Omega_c}{2}(\tilde{\rho}_{12}-\tilde{\rho}_{21}), \label{eq:cwdrho22}\\
  \partial_t\tilde{\rho}_{10} &=& -\Gamma_1\tilde{\rho}_{10}-\frac{i\Omega_p}{4}(\tilde{\rho}_{11}-\tilde{\rho}_{00})
  +\frac{i\Omega_c}{2}\tilde{\rho}_{20}, \label{eq:cwdrho10}\\
  \partial_t\tilde{\rho}_{20} &=& -\Gamma_2\tilde{\rho}_{20}-\frac{i\Omega_p}{4}\tilde{\rho}_{21}
  +\frac{i\Omega_c}{2}\tilde{\rho}_{10}, \label{eq:cwdrho20}\\
  \partial_t\tilde{\rho}_{21} &=& -\Gamma\tilde{\rho}_{21}+\frac{i\Omega_c}{2}(\rho_{11}-\rho_{22})-\frac{i\Omega_p}{4}
  \tilde{\rho}_{20}\label{eq:cwdrho21},
\end{eqnarray}
with $\tilde{\rho}_{01}=\tilde{\rho}_{01}^*$, $\tilde{\rho}_{12}=\tilde{\rho}_{21}^*$, $\tilde{\rho}_{02}=\tilde{\rho}_{20}^*$,
$\Gamma_2=\gamma_{21}/2+\gamma_{22}$ and $\Gamma=\Gamma_1+\Gamma_2$. As the atom is initially in the ground level $|0\rangle$,
$\tilde{\rho}_{00}(0)=1$, $\tilde{\rho}_{11}(0)=\tilde{\rho}_{22}(0)=0$ and $\tilde{\rho}_{21}(0)=0$. Combined with
$\tilde{\rho}_{00}+\tilde{\rho}_{11}+\tilde{\rho}_{22}=1$ and weak probe field condition ($\Omega_p\ll\Omega_c$), the first-order steady state solutions to Eqs.~(\ref{eq:cwdrho00})-(\ref{eq:cwdrho21}) are
\begin{eqnarray}\label{eq:cwrho10}
  \tilde{\rho}_{10} &=& \frac{i\Omega_p\Gamma_2}{4\Gamma_1\Gamma_2+\Omega_c^2},~~
  \tilde{\rho}_{21} = \frac{-i\Omega_c\Omega_p^2/(32\Gamma)}{\Gamma_1\Gamma_2+\Omega_c^2/4}, \\
  \tilde{\rho}_{20} &=& \frac{-\Omega_c\Omega_p/8}{\Gamma_1\Gamma_2+\Omega_c^2/4},~
  \tilde{\rho}_{11}= \frac{\Omega_p^2\Gamma_2}{8\gamma_{10}(\Gamma_1\Gamma_2+\Omega_c^2/4)}, \\
  \tilde{\rho}_{22}&=& \frac{\Omega_p^2\Omega_c^2}{32\gamma_{21}\Gamma(\Gamma_1\Gamma_2+\Omega_c^2/4)}\label{eq:cwrho22}.
  %\tilde{\rho}_{00}&=&1-\tilde{\rho}_{11}-\tilde{\rho}_{22}
\end{eqnarray}

After straightforward calculations, one obtains the micromotion part $\sigma_{\text{MM}}(t)$ and then the off-diagonal
 term $\rho_{\text{MM}}^{10}(t)$:
\begin{eqnarray}
  \rho_{\text{MM}}^{10}(t)&=&
  \frac{-i \Omega _p}{\pi  \omega } \left(1-\frac{\left(8 \Gamma  \Gamma _2 \gamma _{21}+\gamma _{10} \Omega _c^2\right) \Omega _p^2}
  {8\text{   }\Gamma  \gamma _{10} \gamma _{21} \left(4 \Gamma_1\Gamma_2+\Omega_c^2\right)}\right)
  \sum_{n=1}^{\infty}\frac{\cos[(2n-1)\omega t]}{(2n-1)^2},
\end{eqnarray}
which is similar to Eq.~(\ref{eq:microt}) and the higher order terms than $O(\omega^{-1})$ are omitted, except for the amplitude of the periodic triangular pulse. It is convenient to
give the NESS of $\rho_{10}(t)$:
\begin{eqnarray}\label{eq:nesscw}
 \rho_{10}^{\text{ness}}(t)&=&\frac{i\Omega_p\Gamma_2}{4\Gamma_1\Gamma_2+\Omega_c^2}-
 \left(1-\frac{\left(8 \Gamma  \Gamma _2 \gamma _{21}+\gamma _{10} \Omega _c^2\right) \Omega _p^2}{8\text{   }\Gamma  \gamma _{10}\gamma _{21} \left(4 \Gamma_1\Gamma_2+\Omega_c^2\right)}\right)
 \frac{i \Omega _p}{\pi  \omega }\sum_{n=1}^{\infty}\frac{\cos[(2n-1)\omega t]}{(2n-1)^2}.
 \end{eqnarray}

In Fig.~\ref{fig:cw}, we present the numerical and analytical results of the off-diagonal element $\text{Im}(\rho_{10})$ for $\Omega_c/(2\pi)=50$, which is larger than $\Omega_p/(2\pi)=0.5$ to satisfy the valid condition of Eqs.~(\ref{eq:cwrho10})-(\ref{eq:cwrho22}) but smaller than Floquet frequency $\omega/(2\pi)=100$ to satisfy the condition of the high-frequency expansions. From the figure, we find that the system reaches NESS after a long time, in which $\text{Im}(\rho_{10}^{\text{ness}})$ periodically oscillates in the form of a triangular wave with an average value close to zero, as predicted by Eq.~(\ref{eq:nesscw}). Moreover, we find from the figure that sometimes $\text{Im}(\rho_{10}^{\text{ness}})$ is far away from zero, because of the large amplitude of the triangular wave. So the reflection coefficient ($r\propto\text{Im}\rho_{10}^{\text{ness}}$) of the input pulse controlled by the CW field is also time-dependent, and sometimes far away from zero, which is not a valid ON state for switching. Therefore, the CW control field isn't a practical switching for the pulsed input field. In the following, we alternatively study the properties of the switching with an SW control field.

\subsubsection{Square-wave (SW) control field}\label{sec:swth}
\begin{figure}[thp]
\centering
\includegraphics[width=3.2in]{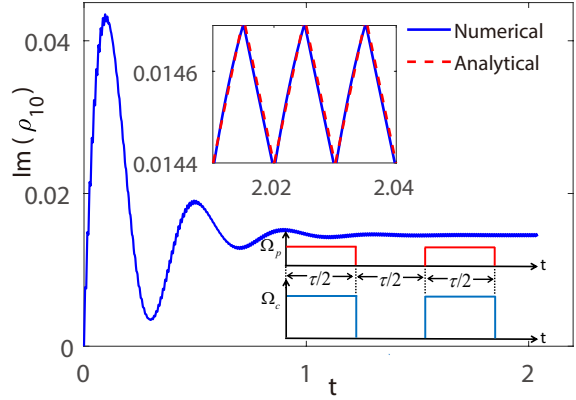}
\caption{\label{fig:sw} Time evolutions of the off-diagonal element $\text{Im}(\rho_{10})$ for $\tau=0.01$,
$\Omega_p/(2\pi)=0.5$, $\Omega_c/(2\pi)=10$, and $\Delta=0$. The blue line denotes the numerical result calculated from the Floquet-Lindblad equation [Eq.~(\ref{eq:tma})] with time-dependent Hamiltonian Eq.~(\ref{eq:threesw}), a sufficiently short time step (here, $\delta t=0.00001$) and an initial state $|0\rangle$. Lower inset: the amplitude envelops of the input and control fields. Upper inset: comparison of the numerical and analytical results of $\text{Im}(\rho_{10})$, when the system reaches NESS. The red dashed line is the analytical result calculated from Eq.~(\ref{eq:swmm})+$\text{Im}(\tilde{\rho}_{10})$. Clearly, the analytical and numerical results agree well with each other.}
\end{figure}
Assuming the SW envelope of strong control field has the same frequency and initial phase with that of the input field, as the lower inset shown in Fig.~\ref{fig:sw}. The Hamiltonian is
 \begin{eqnarray}\label{eq:threesw}
  H(t) &=& \frac{\Delta}{2}(|1\rangle\langle1|+|2\rangle\langle2|-|0\rangle\langle0|)
   - \frac{1}{2}[\Omega_p(t)|0\rangle\langle1|+\Omega_c(t)|1\rangle\langle2|+h.c.],
\end{eqnarray}
where
\begin{eqnarray}\label{eq:omep}
\Omega_{c}(t)&=&\frac{\Omega_p}{2}+\sum_{n=1}^{\infty}\Omega_{cn}\sin(\omega_n t),\\
\Omega_{cn} &=& \frac{2\Omega_c}{(2n-1)\pi}, \\
\omega_n &=& (2n-1)\omega,~~~n=1,2,3\cdots.
\end{eqnarray}
Similarly, the Floquet matrix blocks are
\begin{eqnarray}
  H_0 &=& \frac{\Delta}{2}(|2\rangle\langle2|+|1\rangle\langle1|-|0\rangle\langle0|)
  -[\frac{\Omega_p}{4}|0\rangle\langle1|+\frac{\Omega_c}{4}|1\rangle\langle2|+h.c.], \label{eq:swh0}\\
  H_{2n-1} &=&H^*_{-(2n-1)}=\frac{i\Omega_{pn}}{4}|0\rangle\langle1|+\frac{i\Omega_{pn}}{4}|1\rangle\langle0|
  +\frac{i\Omega_{cn}}{4}|1\rangle\langle2|+\frac{i\Omega_{cn}}{4}|2\rangle\langle1|, \\
  H_{2n} &=& H^*_{-2n}=0.
\end{eqnarray}
Note that the time-dependent parts of H (i.e., $H_{n\neq0}$) include large elements $\Omega_{cn}$, which are beyond the
scope of the high-frequency expansion formula shown in Sec.~\ref{sec:form}. However, for weak control field (such as, $\Omega_c\ll\omega$), the formula is valid. Omitting the higher order terms than $O(\omega^{-1})$, the micromotion part $\sigma_{\text{MM}}(t)$ of $\rho_{10}$ is
\begin{eqnarray}\label{eq:swmm}
  \rho_{\text{MM}}^{10}(t)&=&\frac{i[(\tilde{\rho}_{11}-
  \tilde{\rho}_{00})\Omega_p-\Omega_c\tilde{\rho}_{20}]}{\pi\omega}
  \sum_{n=1}^{\infty}\frac{\cos[(2n-1)\omega t]}{(2n-1)^2}.
\end{eqnarray}
Note that here the results of $\tilde{\rho}$ cannot be simply given by replacing $\Omega_c$ in Eqs.~(\ref{eq:cwrho10})-(\ref{eq:cwrho22}) with $\Omega_c/2$, because Eqs.~(\ref{eq:cwrho10})-(\ref{eq:cwrho22})
is established under the condition that $\Omega_c\gg\Omega_p$. However, the numerical results of $\tilde{\rho}$ for any $\Omega_c$ can be obtained by calculating Eq.~(\ref{eq:rhot}) with $H_0$ in Eq.~(\ref{eq:swh0}).

In Fig.~\ref{fig:sw}, we present the numerical and analytical results obtained from Eq.~(\ref{eq:swmm}) for $\Omega_c/(2\pi)=10$. From the figure, we find that the system reaches NESS after a long time, and with the SW control field on, the reflection coefficient of the input field is also time-dependent but the oscillation amplitude is very small, which indicates that the SW control field may be a practical switching for the pulsed input field. To explore this problem more comprehensively, in the following, we further study the properties of the switching controlled by a CW/SW field with a wide range of intensity.

\section{All-optical pulse switching}\label{sec:allo}

In the above section, we have obtained the analytical expressions of the NESSs of the periodically driven dissipative systems under special conditions. These analytical results help us to better understand the NESS of the periodically driven dissipative system, which is the key issue in designing AOPS. For the dissipative two-level system with the control filed off, Eq.~(\ref{eq:nesstwo}) gives the analytical results in a wide range of parameters.  However, for the dissipative three-level systems with SW control field, the time-dependent parts of H (i.e., $H_{\text{n}\neq0}$) include large elements $\Omega_{\text{cn}}$, which break down the condition of the formula in Sec.~\ref{sec:form}. In this section, to comprehensively investigate the properties of the AOPS in a wide range of $\Omega_c$, we numerically integrate the Floquet-Lindblad equation [Eq.~(\ref{eq:tma})] with a short time step (i.e., $\delta t\ll \tau$) until the systems reach the NESSs.

\subsection{All-optical pulse switching with a strong CW control field vs an SW control field}

\begin{figure}[thp]
\centering
\includegraphics[width=3.2in]{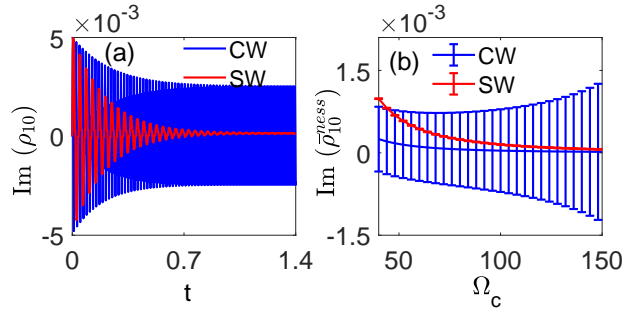}
\caption{\label{fig:num} (a) Numerical results of the time evolutions of the off-diagonal element
$\text{Im}(\rho_{10})$ with a CW control field (blue line) on and an SW control field (red line) on. The parameters
are $\tau=0.01$, $\Omega_p/(2\pi)=0.5$, $\Omega_c/(2\pi)=100$, $\Delta=0$ and time step $\delta t=0.00001$. After a
sufficiently long time, the system reaches NESS. (b) The average and standard deviation (error bar) of $\text{Im}(\rho_{10}^{\text{ness}})$ as a function of $\Omega_c$ for CW control field (blue line) and SW control field (red line). Here the total length of the error bar is the standard deviation.}
\end{figure}

\begin{figure}[thp]
\centering
\includegraphics[width=3.2in]{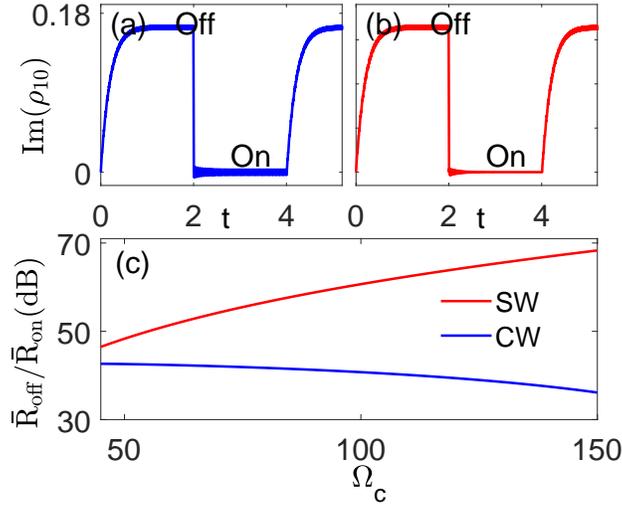}
\caption{\label{fig:offon} Numerical simulations of $\text{Im}(\rho_{10})$ for a switching event with
a CW control field (a) and an SW control field (b). The parameters are $\tau=0.01$, $\Omega_p/(2\pi)=0.5$, $\Omega_c/(2\pi)=120$, $\Delta=0$ and time step $\delta t=0.00001$. (c) The OFF/ON power ratio $\bar{R}_{\text{off}}/\bar{R}_{\text{on}}$ of the switchings controlled by a CW field (blue line) and an SW field (red line) as a function of $\Omega_c$. Obviously, within the parameter range shown in the figure, the $\bar{R}_{\text{off}}/\bar{R}_{\text{on}}$ value of the switching controlled by the SW field is larger than that of the switching controlled by the CW field.}
\end{figure}

Switching has two states: (i) OFF state, where the input field is reflected by the system with control field off and (ii) ON state, where the input field transmits through the system with a strong control field on, i.e., the reflection coefficient $r\rightarrow0$.  For the OFF state, the reflection coefficient $r\propto\text{Im}(\rho_{10}^{\text{ness}})$ is given by Eq.~(\ref{eq:nesstwo}). For the ON state, in Fig.~\ref{fig:num}(a) we compare the time evolutions of $\text{Im}(\rho_{10})$ with a strong CW control field (blue line) on and a strong SW control field (red line) on. Both of them oscillate for a sufficiently long time and then reach NESSs. After reaching NESSs, the oscillation amplitude of $\text{Im}(\rho_{10})$ controlled by the CW field is much larger than that controlled by SW field, which denotes that the reflection coefficient $r$ controlled by CW field oscillates violently and sometimes far away from zero. Moreover, in Fig.~\ref{fig:num}(b), we present the average and the standard deviation of $\text{Im}(\rho_{10}^{\text{ness}})$ controlled by a CW field (blue line) and an SW field (red line) as a function of $\Omega_c$. We find that although their averages of $\text{Im}(\rho_{10}^{\text{ness}})$ are very close and approximate to 0, the standard deviations of $\text{Im}(\rho_{10}^{\text{ness}})$ are quite different and the latter is smaller. Therefore, the SW field is more suitable to switch the pulsed input field.

In Fig.~\ref{fig:offon}, we present the results of time-domain numerical simulations of $\text{Im}(\rho_{10})$ for a switching event corresponding to the one occurring in the AOPS controlled by a CW field [Fig.~\ref{fig:offon}(a)] and an SW field [Fig.~\ref{fig:offon}(b)]. One sees that the transients tend to NESSs and the oscillation amplitude of the ON state in Fig.~\ref{fig:offon}(a) is larger than that in Fig.~\ref{fig:offon}(b). Note that here the SW period is much larger than the rising/falling edge of the SW, and much smaller than the decoherence time of the system, i.e., $\tau=0.01\ll 1/\Gamma\approx0.67$ with total decay rate $\Gamma=(\gamma_{10}+\gamma_{21})/2+\gamma_{11}+\gamma_{22}=1.5$, resulting in at least 67 SW periods in a single switching time (i.e., the
reaction time of the system to reach NESS). Therefore, when the SW control field is turned on/off within an SW period has a negligible effect on the switching effect. Additionally, the switching time is independent of whether the control field is CW or SW and depends on the decay rate of the system. Moreover, as a factor of merit for switching, the OFF/ON ratio between the reflected powers ($\text{R}=|r|^2$) in the OFF state and that in the ON state $\text{R}_{\text{off}}/\text{R}_{\text{on}}$ must be as large as possible. According to Eq.~(\ref{eq:reco}), $\text{R}_{\text{off}}/\text{R}_{\text{on}}=|\text{Im}(\rho_{10}^{\text{ness}})_{\text{off}}|^2/|\text{Im}(\rho_{10}^{\text{ness}})_{\text{on}}|^2$.
Since the reflection coefficients in NESSs also oscillate with time, we calculate the average of reflected powers
$\bar{\text{R}}_{\text{off}}$ and $\bar{\text{R}}_{\text{on}}$. In Fig.~\ref{fig:offon}(c), we further present the
$\bar{\text{R}}_{\text{off}}/\bar{\text{R}}_{\text{on}}$ ratio in the unit of $\text{dB}$ as a function of $\Omega_c$. It is clear that as the intensity of the control field increases, the values of $\bar{\text{R}}_{\text{off}}/\bar{\text{R}}_{\text{on}}$ controlled by the SW field increase, while the ones controlled by the CW field decreases. Moreover, the value of $\bar{\text{R}}_{\text{off}}/\bar{\text{R}}_{\text{on}}$ controlled by the SW field can be as high as nearly 70dB, which is much higher than that of the switching (about 50dB) in Ref.~\cite{A2018An}. Obviously, for a high-performance switching, the strong SW control field is more advantageous. In the following section, we will explore the robustness of $\bar{\text{R}}_{\text{off}}/\bar{\text{R}}_{\text{on}}$ to the pulse errors.

\subsection{Robustness of the All-optical pulse switching with a strong SW control field}
\begin{figure}[thp]
\centering
\includegraphics[width=3.2in]{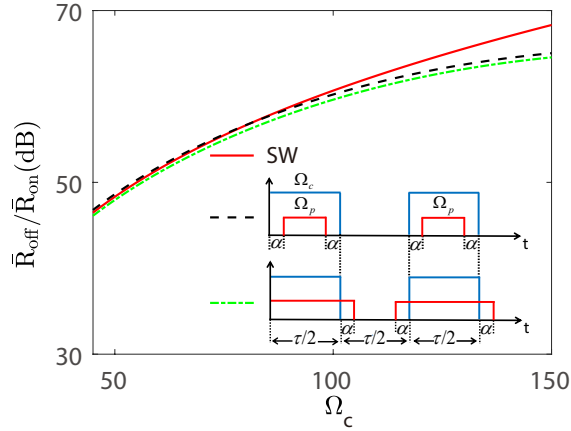}
\caption{\label{fig:robust} Robustness of $\bar{\text{R}}_{\text{off}}/\bar{\text{R}}_{\text{on}}$ against
 the mismatch pulses of the control and input fields. The parameters are the same as that in Fig.~\ref{fig:offon}(c)
 with $\alpha=0.01\tau$. The inset shows the pulse sequences corresponding to the black dashed line and the green dot-dashed line. The matched SW control (red solid line) results are also presented for comparisons.}
\end{figure}

To estimate the robustness of $\bar{\text{R}}_{\text{off}}/\bar{\text{R}}_{\text{on}}$ to the pulse errors, we consider
mismatch pulses between the control and the input fields, such as, assuming a nonoverlap time $\alpha$ between the SW envelopes of the input and control fields, as the inset pulse sequences shown in Fig.~\ref{fig:robust}. The black dashed line denotes that the opening time of the control field in one cycle completely covers the signal time of the input field and is redundant. On the contrary, the green dot-dashed line denotes that the time that the control field is turned on in one cycle does not completely cover the signal time of the input field. Remarkably, the OFF/ON powers ratios $\bar{\text{R}}_{\text{off}}/\bar{\text{R}}_{\text{on}}$ are quite close and show strong robustness against the mismatch pulses between control and the input fields, especially when the strength of the control field is in the range $\Omega_c \in [50, 100]$.

\section{Conclusion}\label{sec:con}
In conclusion, we propose an AOPS in a dissipative three-level system, in which an effective SW control field is used to switch an input field whose amplitude envelope is a periodic SW pulse. By comparing the NESSs of the dissipative systems driven by a CW and an SW field from analytical and numerical results, we find that the strong SW field with the same parameters as the input field (except for the Rabi-frequency), is a practical field for switching the input field and show strong robustness against the mismatch pulses between control and the input fields. Our theoretical protocol may be generalized to other pulse forms of the input field. This work may find potential applications in quantum computation processing and quantum networks based on all-optical devices.

\section{Appendix: An all-optical pulse switching with electromagnetically induced transparency}\label{sec:app}
Here, the all-optical pulse switching scheme is demonstrated with electromagnetically induced transparency parameters, i.e., same as the ones in the main text except for $\gamma_{21}/(2\pi)=0.1$ and $\gamma_{22}/(2\pi)=0.01$, which satisfy the condition that a transparent window is induced by destructive interference~\cite{2016Interference}, as shown in Fig.~\ref{fig:EIT} (a). Same as the main text, for an SW-pulse input field, in Fig.~\ref{fig:EIT} (b) we compare the time evolutions of $\text{Im}(\rho_{10})$ with a strong CW control field (blue line) on and a strong SW control field (red line) on. Clearly, after reaching NESSs, the oscillation amplitude of $\text{Im}(\rho_{10})$ controlled by the CW field is much larger than that controlled by SW field. Moreover, in Fig.~\ref{fig:EIT} (c), we present the average and the standard deviation of $\text{Im}(\rho_{10}^{\text{ness}})$ controlled by a CW field (blue line) and an SW field (red line) as a function of $\Omega_c$. We find that the standard deviations of $\text{Im}(\rho_{10}^{\text{ness}})$ are quite different and the latter is much smaller. In Fig.~\ref{fig:EIT} (d), we further present the
$\bar{\text{R}}_{\text{off}}/\bar{\text{R}}_{\text{on}}$ ratio in the unit of $\text{dB}$ as a function of $\Omega_c$. It is clear that the SW field is more suitable to switch the pulsed input field. Therefore, our all-optical pulsing scheme is also suitable to the electromagnetically induced transparency.

\begin{figure}[htbp]
\centering
\includegraphics[width=.6\linewidth]{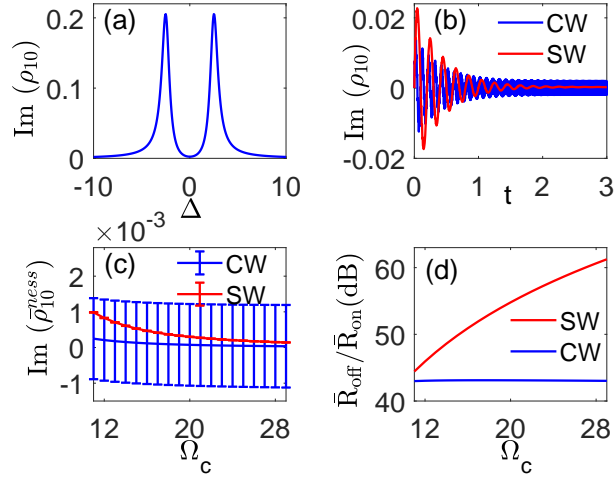}
\caption{(a) Electromagnetically induced transparency window with $\Omega_c/(2\pi)=5$. (b) Numerical results of the time evolutions of the off-diagonal element
$\text{Im}(\rho_{10})$ with a CW control field (blue line) on and an SW control field (red line) on. The parameters
are  $\Omega_c/(2\pi)=20$, $\Delta=0$ and time step $\delta t=0.00001$. After a
sufficiently long time, the system reaches NESS. (c) The average and standard deviation (error bar) of $\text{Im}(\rho_{10}^{\text{ness}})$ as a function of $\Omega_c$ for the CW control field (blue line) and the SW control field (red line) with $\Delta=0$. Here the size of the error bar is the standard deviation. (d) The OFF/ON power ratio $\bar{R}_{\text{off}}/\bar{R}_{\text{on}}$ of the switchings controlled by a CW field (blue line) and an SW field (red line) as a function of $\Omega_c$ with $\Delta=0$. Obviously, within the parameter range shown in the figure, the $\bar{R}_{\text{off}}/\bar{R}_{\text{on}}$ value of the switching controlled by the SW field is larger than that of the switching controlled by the CW field. Note that the value of $\Omega_c$ starts from 11, because the intensity of the same electromagnetic field is halved under SW modulation. Other parameters are shown in Sec.~\ref{sec:app}.}
\label{fig:EIT}
\end{figure}

\begin{backmatter}
\bmsection{Funding} This work is supported by the National Natural Science Foundation of China (Grant No. 11874247 and No. U1930201), the National Key Research and Development Program of China (Grant Nos. 2017YFA0304500 and 2017YFA0304203), the Guangdong Provincial Key Laboratory (Grant No. 2019B121203002).

\bmsection{Acknowledgments} We thank Xiuhao Deng for many helpful and intriguing discussions.

\bmsection{Disclosures} The authors declare no conflicts of interest.

\bmsection{Data Availability Statement} Data underlying the results presented in this paper are not publicly available at this time but may be obtained from the authors upon reasonable request.
\end{backmatter}

%%%%%%%%%%%%%%%%%%%%%%% References %%%%%%%%%%%%%%%%%%%%%%%%%

%%%%%%%%%% If using BibTeX:
\bibliography{ref}

\begin{thebibliography}{10}
\newcommand{\enquote}[1]{``#1''}

\bibitem{Minzioni_2019}
P.~Minzioni, C.~Lacava, T.~Tanabe, J.~Dong, X.~Hu, G.~Csaba, W.~Porod,
  G.~Singh, A.~E. Willner, A.~Almaiman, V.~Torres-Company, J.~Schröder, A.~C.
  Peacock, M.~J. Strain, F.~Parmigiani, G.~Contestabile, D.~Marpaung, Z.~Liu,
  J.~E. Bowers, L.~Chang, S.~Fabbri, M.~R. V{\'{a}}zquez, V.~Bharadwaj, S.~M.
  Eaton, P.~Lodahl, X.~Zhang, B.~J. Eggleton, W.~J. Munro, K.~Nemoto, O.~Morin,
  J.~Laurat, and J.~Nunn, \enquote{Roadmap on all-optical processing,}
  {\protect\JournalTitle{J. Opt.}} \textbf{21}, 063001 (2019).

\bibitem{MXene2020}
W.~U. Qing, Y.~Wang, W.~Huang, C.~Wang, Z.~Zheng, M.~Zhang, and H.~Zhang,
  \enquote{Mxene-based high-performance all-optical modulators for actively
  q-switched pulse generation,} {\protect\JournalTitle{Photonics Res.}}
  \textbf{8}, 1140--1147 (2020).

\bibitem{Yuan:21}
J.~Yuan, S.~Dong, H.~Zhang, C.~Wu, L.~Wang, L.~Xiao, and S.~Jia,
  \enquote{Efficient all-optical modulator based on a periodic dielectric
  atomic lattice,} {\protect\JournalTitle{Opt. Express}} \textbf{29},
  2712--2719 (2021).

\bibitem{An2010}
L.~Liu, R.~Kumar, K.~Huybrechts, T.~Spuesens, G.~Roelkens, E.~J. Geluk, T.~D.
  Vries, P.~Regreny, D.~V. Thourhout, and R.~Baets, \enquote{An ultra-small,
  low-power, all-optical flip-flop memory on a silicon chip,}
  {\protect\JournalTitle{Nat. Photonics}} \textbf{4}, 182--187 (2010).

\bibitem{IST2005}
F.~Ramos, E.~Kehayas, J.~M. Martinez, R.~Clavero, and B.~Riposati,
  \enquote{Ist-lasagne: Towards all-optical label swapping employing optical
  logic gates and optical flip-flops,} {\protect\JournalTitle{J. Lightwave
  Technol.}} \textbf{23}, 2993--3011 (2005).

\bibitem{Optical2010}
A.~M. Kaplan, G.~P. Agrawal, and D.~N. Maywar, \enquote{Optical square-wave
  clock generation based on an all-optical flip-flop,}
  {\protect\JournalTitle{IEEE Photonics Technol. Lett.}} \textbf{22}, 489--491
  (2010).

\bibitem{Ultrafast}
T.~Volz, A.~Reinhard, M.~Winger, A.~Badolato, K.~J. Hennessy, E.~L. Hu, and
  A.~Imamolu, \enquote{Ultrafast all-optical switching by single photons,}
  {\protect\JournalTitle{Nat. Photonics}} \textbf{6}, 605--609 (2012).

\bibitem{All2012}
Y.~Fu, X.~Hu, C.~Lu, S.~Yue, H.~Yang, and Q.~Gong, \enquote{All-optical logic
  gates based on nanoscale plasmonic slot waveguides.}
  {\protect\JournalTitle{Nano Lett.}} \textbf{12}, 5784--5790 (2012).

\bibitem{Optical2005}
Z.~Li, Z.~Chen, and B.~Li, \enquote{Optical pulse controlled all-optical logic
  gates in sige/si multimode interference,} {\protect\JournalTitle{Opt.
  Express}} \textbf{13}, 1033--1038 (2005).

\bibitem{All2006}
A.~Betlej, S.~Suntsov, K.~G. Makris, L.~Jankovic, D.~N. Christodoulides, G.~I.
  Stegeman, J.~Fini, R.~T. Bise, and D.~J. Digiovanni, \enquote{All-optical
  switching and multifrequency generation in a dual-core photonic crystal
  fiber,} {\protect\JournalTitle{Opt. Lett.}} \textbf{31}, 1480--1482 (2006).

\bibitem{Almeida}
V.~R. Almeida, C.~A. Barrios, R.~R. Panepucci, M.~Lipson, M.~A. Foster, D.~G.
  Ouzounov, and A.~L. Gaeta, \enquote{All-optical switching on a silicon chip,}
  {\protect\JournalTitle{Opt. Lett.}} \textbf{29}, 2867--2869 (2004).

\bibitem{1992All}
N.~D. Sankey, D.~F. Prelewitz, and T.~G. Brown, \enquote{All‐optical
  switching in a nonlinear periodic‐waveguide structure,}
  {\protect\JournalTitle{Appl. Phys. Lett.}} \textbf{60}, 1427--1429 (1992).

\bibitem{Boyraz}
\"{O}zdal Boyraz, P.~Koonath, V.~Raghunathan, and B.~Jalali, \enquote{All
  optical switching and continuum generation in silicon waveguides,}
  {\protect\JournalTitle{Opt. Express}} \textbf{12}, 4094--4102 (2004).

\bibitem{LiDynamical}
J.~Li, G.~S. Paraoanu, K.~Cicak, F.~Altomare, J.~I. Park, R.~W. Simmonds, M.~A.
  Sillanp\"a\"a, and P.~J. Hakonen, \enquote{Dynamical autler-townes control of
  a phase qubit,} {\protect\JournalTitle{Sci. Rep.}} \textbf{2}, 645 (2012).

\bibitem{Shaped}
R.~Bartels, S.~Backus, E.~Zeek, L.~Misoguti, G.~Vdovin, I.~P. Christov, M.~M.
  Murnane, and H.~C. Kapteyn, \enquote{Shaped-pulse optimization of coherent
  emission of high-harmonic soft x-rays,} {\protect\JournalTitle{Nature}}
  \textbf{406}, 164--166 (2000).

\bibitem{Malinovskaya}
S.~A. Malinovskaya and V.~S. Malinovsky, \enquote{Chirped-pulse adiabatic
  control in coherent anti-stokes raman scattering for imaging of biological
  structure and dynamics,} {\protect\JournalTitle{Opt. Lett.}} \textbf{32},
  707--709 (2007).

\bibitem{Parmigiani}
F.~Parmigiani, P.~Petropoulos, M.~Ibsen, and D.~J. Richardson,
  \enquote{All-optical pulse reshaping and retiming systems incorporating pulse
  shaping fiber bragg grating,} {\protect\JournalTitle{J. Lightwave Technol.}}
  \textbf{24}, 357--364 (2006).

\bibitem{Ultrafast2007}
Y.~Ukai, N.~Nishizawa, and T.~Goto, \enquote{Ultrafast all-optical switching
  using pulse trapping by ultrashort soliton pulse in birefringent optical
  fiber,} {\protect\JournalTitle{Wiley Subscription Services, Inc., A Wiley
  Company}} \textbf{158}, 38--44 (2007).

\bibitem{PhysRevLett.68.911}
M.~E. Crenshaw, M.~Scalora, and C.~M. Bowden, \enquote{Ultrafast intrinsic
  optical switching in a dense medium of two-level atoms,}
  {\protect\JournalTitle{Phys. Rev. Lett.}} \textbf{68}, 911--914 (1992).

\bibitem{Ramos}
P.~Ramos and C.~Paiva, \enquote{All-optical pulse switching in twin-core fiber
  couplers with intermodal dispersion,} {\protect\JournalTitle{IEEE J. Quantum
  Electron.}} \textbf{35}, 983--989 (1999).

\bibitem{PhysRevLett.95.043001}
R.~Garcia-Fernandez, A.~Ekers, L.~P. Yatsenko, N.~V. Vitanov, and K.~Bergmann,
  \enquote{Control of population flow in coherently driven quantum ladders,}
  {\protect\JournalTitle{Phys. Rev. Lett.}} \textbf{95}, 043001 (2005).

\bibitem{NOVITSKY2016202}
D.~V. Novitsky, \enquote{Compression and collisions of chirped pulses in a
  dense two-level medium,} {\protect\JournalTitle{Opt. Commun.}} \textbf{358},
  202--207 (2016).

\bibitem{BirnbaumPhoton}
K.~Birnbaum, A.~Boca, R.~Miller, A.~Boozer, T.~Northup, and H.~Kimble,
  \enquote{Photon blockade in an optical cavity with one trapped atom,}
  {\protect\JournalTitle{Nature}} \textbf{436}, 87--90 (2005).

\bibitem{Dawes2005All}
A.~M. Dawes, L.~Illing, S.~M. Clark, and D.~J. Gauthier, \enquote{All-optical
  switching in rubidium vapor,} {\protect\JournalTitle{Science}} \textbf{308},
  672--674 (2005).

\bibitem{PhysRevA.68.041801}
D.~A. Braje, V.~Bali\ifmmode~\acute{c}\else \'{c}\fi{}, G.~Y. Yin, and S.~E.
  Harris, \enquote{Low-light-level nonlinear optics with slow light,}
  {\protect\JournalTitle{Phys. Rev. A}} \textbf{68}, 041801 (2003).

\bibitem{Wang2002Controlling}
H.~Wang, D.~Goorskey, and M.~Xiao, \enquote{Controlling light by light with
  three-level atoms inside an optical cavity,} {\protect\JournalTitle{Opt.
  Lett.}} \textbf{27}, 1354--1356 (2002).

\bibitem{Chen:05}
Y.-F. Chen, Z.-H. Tsai, Y.-C. Liu, and I.~A. Yu, \enquote{Low-light-level
  photon switching by quantum interference,} {\protect\JournalTitle{Opt.
  Lett.}} \textbf{30}, 3207--3209 (2005).

\bibitem{PhysRevLett.102.203902}
M.~Bajcsy, S.~Hofferberth, V.~Balic, T.~Peyronel, M.~Hafezi, A.~S. Zibrov,
  V.~Vuletic, and M.~D. Lukin, \enquote{Efficient all-optical switching using
  slow light within a hollow fiber,} {\protect\JournalTitle{Phys. Rev. Lett.}}
  \textbf{102}, 203902 (2009).

\bibitem{Stern2017Strong}
L.~Stern, B.~Desiatov, N.~Mazurski, and U.~Levy, \enquote{Strong coupling and
  high-contrast all-optical modulation in atomic cladding waveguides.}
  {\protect\JournalTitle{Nat. Commun.}} \textbf{8}, 14461 (2017).

\bibitem{PhysRevB.104.165414}
A.~Schnell, S.~Denisov, and A.~Eckardt, \enquote{High-frequency expansions for
  time-periodic lindblad generators,} {\protect\JournalTitle{Phys. Rev. B}}
  \textbf{104}, 165414 (2021).

\bibitem{2020General}
T.~N. Ikeda and M.~Sato, \enquote{General description for nonequilibrium steady
  states in periodically driven dissipative quantum systems,}
  {\protect\JournalTitle{Sci. Adv.}} \textbf{6}, 4019 (2020).

\bibitem{PhysRevA.101.022108}
Y.~Han, X.-Q. Luo, T.-F. Li, and W.~Zhang, \enquote{Analytical
  double-unitary-transformation approach for strongly and periodically driven
  three-level systems,} {\protect\JournalTitle{Phys. Rev. A}} \textbf{101},
  022108 (2020).

\bibitem{PhysRevLett.95.213001}
J.-T. Shen and S.~Fan, \enquote{Coherent single photon transport in a
  one-dimensional waveguide coupled with superconducting quantum bits,}
  {\protect\JournalTitle{Phys. Rev. Lett.}} \textbf{95}, 213001 (2005).

\bibitem{Resonance}
O.~Astafiev, A.~M. Zagoskin, A.~Jr, Y.~A. Pashkin, T.~Yamamoto, K.~Inomata,
  Y.~Nakamura, and J.~S. Tsai, \enquote{Resonance fluorescence of a single
  artificial atom,} {\protect\JournalTitle{Science}} \textbf{327}, 840--843
  (2010).

\bibitem{PhysRevLett.104.193601}
A.~A. Abdumalikov, O.~Astafiev, A.~M. Zagoskin, Y.~A. Pashkin, Y.~Nakamura, and
  J.~S. Tsai, \enquote{Electromagnetically induced transparency on a single
  artificial atom,} {\protect\JournalTitle{Phys. Rev. Lett.}} \textbf{104},
  193601 (2010).

\bibitem{PhysRevLett.107.073601}
I.-C. Hoi, C.~M. Wilson, G.~Johansson, T.~Palomaki, B.~Peropadre, and
  P.~Delsing, \enquote{Demonstration of a single-photon router in the microwave
  regime,} {\protect\JournalTitle{Phys. Rev. Lett.}} \textbf{107}, 073601
  (2011).

\bibitem{2018All}
V.~Sasikala and K.~Chitra, \enquote{All optical switching and associated
  technologies: a review,} {\protect\JournalTitle{J. Opt.}} \textbf{47},
  307–317 (2018).

\bibitem{PhysRevA.64.041801}
M.~Yan, E.~G. Rickey, and Y.~Zhu, \enquote{Observation of absorptive photon
  switching by quantum interference,} {\protect\JournalTitle{Phys. Rev. A}}
  \textbf{64}, 041801 (2001).

\bibitem{PhysRevA.79.032301}
S.-K. Son, S.~Han, and S.-I. Chu, \enquote{{Floquet} formulation for the
  investigation of multiphoton quantum interference in a superconducting qubit
  driven by a strong ac field,} {\protect\JournalTitle{Phys. Rev. A}}
  \textbf{79}, 032301 (2009).

\bibitem{Quantum}
A.~J. Daley, \enquote{Quantum trajectories and open many-body quantum systems,}
  {\protect\JournalTitle{Adv. Phys.}} \textbf{63}, 77--149 (2014).

\bibitem{PhysRevLett.127.070402}
T.~Haga, M.~Nakagawa, R.~Hamazaki, and M.~Ueda, \enquote{Liouvillian skin
  effect: Slowing down of relaxation processes without gap closing,}
  {\protect\JournalTitle{Phys. Rev. Lett.}} \textbf{127}, 070402 (2021).

\bibitem{PhysRevA.93.032121}
C.~M. Dai, Z.~C. Shi, and X.~X. Yi, \enquote{Floquet theorem with open systems
  and its applications,} {\protect\JournalTitle{Phys. Rev. A}} \textbf{93},
  032121 (2016).

\bibitem{HanPRApplied}
Y.~Han, X.-Q. Luo, T.-F. Li, W.~Zhang, S.-P. Wang, J.~S. Tsai, F.~Nori, and
  J.~Q. You, \enquote{Time-domain grating with a periodically driven qutrit,}
  {\protect\JournalTitle{Phys. Rev. Applied}} \textbf{11}, 014053 (2019).

\bibitem{A2018An}
Y.~Wu, L.~P. Yang, M.~Gong, Y.~Zheng, H.~Deng, Z.~Yan, Y.~Zhao, K.~Huang, A.~D.
  Castellano, W.~J. Munro, K.~Nemoto, D.~Zhang, C.~P. Sun, Y.~Liu, and X.~Zhu,
  \enquote{An efficient and compact quantum switch for quantum circuits,}
  {\protect\JournalTitle{npj Quantum Inf.}} \textbf{4}, 50 (2018).

\bibitem{2016Interference}
S.~Davuluri and S.~Zhu, \enquote{Interference via dephasing effect in upper
  coupled three-level atoms,} {\protect\JournalTitle{Phys. Scr.}} \textbf{91},
  013008 (2016).

\end{thebibliography}

\end{document}